%% file: Lietal13TSPsub_Final.tex
\documentclass[journal]{IEEEtran}
\usepackage{mathptmx} 
\usepackage{times} 
\usepackage{amsmath} 
\usepackage{amssymb}  
\usepackage{psfrag}
\usepackage{pmat}
\usepackage[table]{xcolor}
\usepackage{cite}
\usepackage{algorithmic}
\usepackage{array}
\usepackage{stfloats}
\usepackage{graphics} 
\usepackage{graphicx}
\usepackage{subfigure} 
\usepackage{algorithm}
\usepackage{epsfig} 
\usepackage{mathptmx} 
\usepackage{times} 
\usepackage{amsmath} 
\usepackage{amssymb}  
\usepackage{float} 
\usepackage{footnote}
\makesavenoteenv{tabular}
\usepackage{threeparttable}

\newtheorem{theorem}{Theorem}
\newtheorem{corollary}[theorem]{Corollary}

\newtheorem{remark}{Remark}

\usepackage{verbatim,pifont,color,amssymb,amsfonts,amsmath,graphicx,pgf,slashbox,psfrag,ifpdf}
\usepackage[mathscr]{eucal}
\newcommand{\Pmsc}{\mathscr{P}}
\newcommand{\Gmsc}{\mathscr{G}}
\newcommand{\Vmsc}{\mathscr{V}}
\newcommand{\Emsc}{\mathscr{E}}
\newcommand{\Tmsc}{\mathscr{T}}

\def\eg{{\it e.g., \/}}
\def\defeq{\triangleq}
\def\T{\mbox{\tiny T}}
\def\N{\mbox{\tiny N}}
\def\K{\mbox{\tiny K}}
\def\ie{{\it i.e.,\ \/}}

\def\tcr{\textcolor{red}}

\begin{document}
%
\title{Spectral Clustering on Subspace for Parameter Estimation
of Jump Linear Models}

\author{{\large Liang~Li,~\IEEEmembership{Member,~IEEE,}
        Wei~Dong,~\IEEEmembership{Member,~IEEE,}\\
        Yindong~Ji,~\IEEEmembership{Member,~IEEE,}
         and~Lang~Tong~\IEEEmembership{Fellow,~IEEE}}
\thanks{L. Li, W. Dong and Y. Ji are with the Department of Automation, Tsinghua University, and Tsinghua National Laboratory for Information Science and Technology, Beijing 100084, China. (e-mail: liang-li07@mails.tsinghua.edu.cn; \{weidong,jyd\}@mail.tsinghua.edu.cn).}
\thanks{L. Tong is with the School of Electrical and Computer Engineering, Cornell
University, Ithaca, NY 14853, USA (e-mail: ltong@ece.cornell.edu).}
\thanks{This work was supported in part by the Natural Science Foundation of China under Grants 61104019, the National Key Technology Research and Development Program under Grant 2009BAG12A08, the Tsinghua University Initiative Scientific Research Program, and the US National Science Foundation under award CCF 1018115.}
\thanks{Part of this work was submitted to the 2013 IEEE Conf. on Decision and Control, March 2013.}
}
\markboth{Submitted to IEEE Transactions on signal processing,~June~2013}%
{Liang~Li \MakeLowercase{\textit{et al.}}: Spectral Clustering on Subspace for Parameter Estimation
of Jump Linear Models}

\maketitle

\begin{abstract}
 The problem of estimating parameters of a deterministic jump or piecewise linear model is considered.   A subspace technique referred to as spectral clustering on subspace (SCS) algorithm  is proposed to estimate a set of linear model parameters, the model input, and the set of switching epochs.  The SCS algorithm exploits  a block diagonal structure of the system input subspace, which partitions the observation space into separate subspaces, each corresponding to one and only one linear submodel.  A spectral clustering technique is used to label the noisy observations for each submodel, which generates estimates of switching time epoches. A total least squares technique is used to estimate model parameters and the model input.  It is shown that, in the absence of observation noise, the SCS algorithm provides exact parameter identification. At high signal to noise ratios, SCS attains a clairvoyant Cram\'{e}r-Rao bound computed by assuming the labeling of observation samples is perfect.
 \end{abstract}


\vspace{0.5em}

 \ifCLASSOPTIONpeerreview
 \begin{center} \bfseries EDICS Category: 3-BBND \tcr{Check on this.} \end{center}
 \fi
%
\IEEEpeerreviewmaketitle

\section{INTRODUCTION\label{Sec:1}}
\input intro_v5

\section{Problem formulation and assumptions\label{Sec:2}}
\input ProblemFormulation_v4

\section{A Subspace Structure\label{Sec:3}}
\input subspace_v4

\section{Data association via Spectral Clustering\label{Sec:4}}
\input SpectralCluster_v4

\section{The Spectral Clustering Algorithm\label{Sec:5}}
\input SCA_v2

\section{Performance Analysis\label{Sec:6}}
\input Performance_v4


\section{Simulation examples\label{Sec:7}}
\input Simu_v6

\section{Conclusions\label{Sec:8}}
This paper presents a subspace approach to the identification of jump linear or piecewise linear models. The main contribution is the idea of applying spectral clustering on the input subspace, which leads to the SCS algorithm that gives a closed form (in terms of eigenvalues and eigenvectors) identification of system matrices, system input sequence, and switching epochs. SCS exhibits stable numerical behavior in our simulations, thanks to the use of SVD.

It should be noted that the proposed algorithm applies only to the class systems with sufficient number of input-output sensors.  This is a restriction not imposed by some of the other methods. However, when the required rank condition is satisfied, the proposed technique does have some computation and performance advantages.

A number of practical issues are not discussed in the current paper, most significant is the identification of the number $K$ of subsystems within a block of $N$ samples.  In the absence of noise, this is not necessarily difficult as the eigenstructure of $Z$ reveals this information. As a hypothesis testing problem, detecting $K$ without knowing system matrices is nontrivial, nor does the estimating system matrices without knowing $K$. To this end, a joint detection and estimation approach is desirable.


\bibliographystyle{IEEEtran}
\bibliography{IEEEabrv,ref}

\section*{Appendix}
\subsection{Proof of Theorem~1\label{apdex1}}
\begin{IEEEproof}
  The input signal space $DP$ is related to the SVD of $Z$ by some full rank matrix $B$ via $DP=BV^{\T}$. We then have $DD^{\T}=BB^{\T}$. Because $D$ is block diagonal, so is $BB^{\T}$.  Let
  \begin{equation}
    BB^{\T}=diag\left(\Omega_1, \ldots, \Omega_K\right) \triangleq \Omega
  \end{equation}
where $\Omega_i=D_iD_i^{\T}> 0$.  Therefore, $\Omega > 0$.  Normalizeing $B$ and $D$,
  \begin{equation}
    \bar{B}\triangleq \Omega^{-1/2}B,~~\bar{D}\triangleq \Omega^{-1/2}D,
  \end{equation}
  we have
  \begin{equation}
    \bar{B}\bar{B}^{\T}=\bar{B}^{\T}\bar{B}=I,
  \end{equation}
\ie $\bar{B}$ is orthogonal.  From $\bar{D}P=\bar{B}V^{\T}$, we have
  \begin{align}
    VV^{\T}&=P^{\T}\bar{D}^{\T}\bar{D}P\notag\\
    &=P^{\T}\left[\begin{array}{ccc}
               \Lambda_1 &  & \\
               & \ddots & \\
               &  & \Lambda_K
             \end{array}
    \right]P
  \end{align}
  where $\Lambda_i=D_i^{\T}\left(D_iD_i^{\T}\right)^{-1}D_i$.
\end{IEEEproof}

\vspace*{-2\baselineskip}
\begin{IEEEbiography}[{\includegraphics[width=1in,height=1.25in,clip,keepaspectratio]{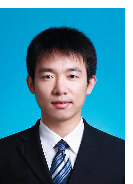}}]{Liang Li}
Liang Li (S'11,M'13) received the B.E. degree in Department of Automation from Xi'an JiaoTong University , Xi'an, China, in 2007. He is currently a Ph.D. candidate in Department of Automation in Tsinghua University. His research interests focus on optimal control and estimation theory of switched and hybrid systems. E-mail: liang-li07@mails.tsinghua.edu.cn.
\end{IEEEbiography}
\vspace*{-2\baselineskip}
\begin{IEEEbiography}[{\includegraphics[width=1in,height=1.25in,clip,keepaspectratio]{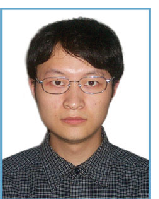}}]{Wei Dong}
(S'03,M'07) received his B.E. and Ph.D. degrees from the Department of Automation, Tsinghua University, Beijing, China, in 2000 and 2006, respectively. He is currently an Assistant Professor with the Department of Automation and the Rail Transit Control Technology Research and Development Center, Tsinghua University. His main research interests include fault diagnosis, modeling and simulation of complex engineering systems. Email: weidong@tsinghua.edu.cn.
\end{IEEEbiography}
\vspace*{-2\baselineskip}
\begin{IEEEbiography}[{\includegraphics[width=1in,height=1.25in,clip,keepaspectratio]{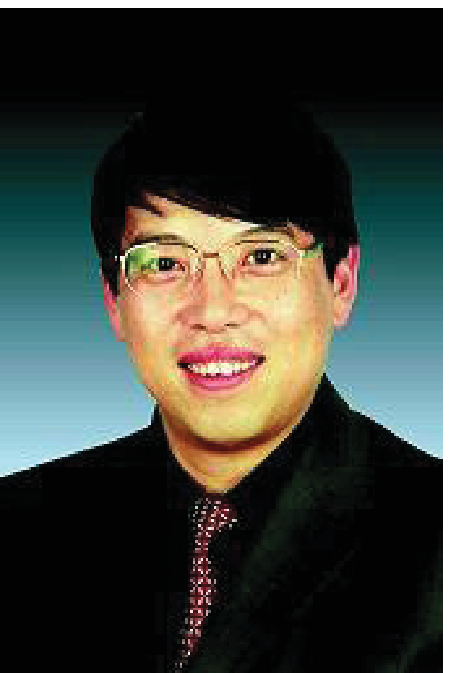}}]{Yingdong Ji}
(M'06) received his B.E. and M.S. degrees from the Department of Automation, Tsinghua University, in 1985 and 1989, respectively. He is currently a Professor with the Department of Automation, the Vice President of the Research Institute of Information Technology, and the Director of the Rail Transit Control Technology Research and Development Center, Tsinghua University. His main research interests include digital signal processing, fault diagnosis and reliability prediction. His current research interests include predictive maintenance and the train control systems of high-speed railways. E-mail: jyd@mail.tsinghua.edu.cn.
\end{IEEEbiography}

\vspace*{-2\baselineskip}

\begin{IEEEbiography}[{\includegraphics[width=1in,height=1.25in,clip,keepaspectratio]{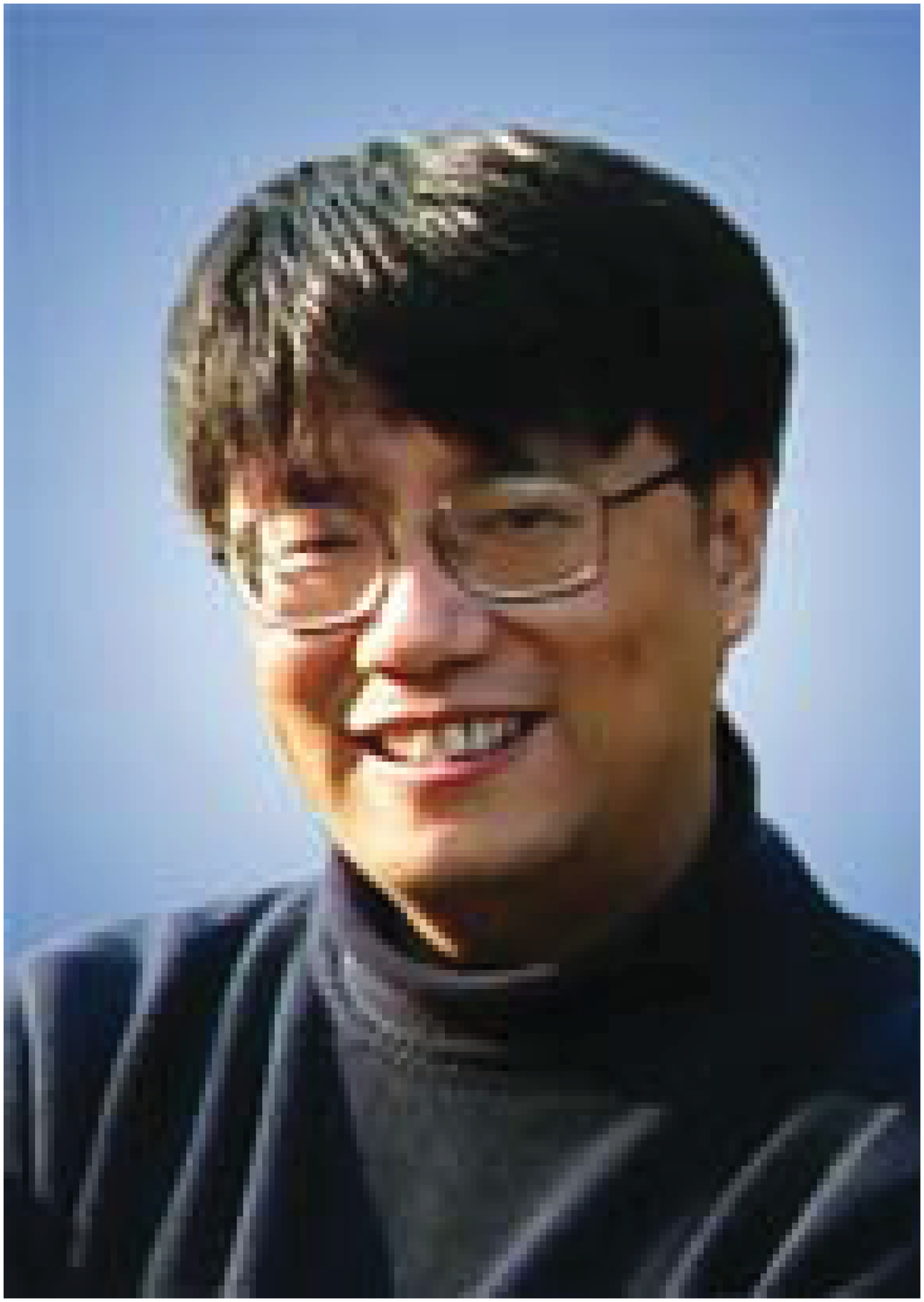}}]{Lang Tong}
(S'87,M'91,SM'01,F'05) is the Irwin and Joan Jacobs Professor in Engineering at Cornell University Ithaca, New York. He received the B.E. degree from Tsinghua University, Beijing, China, in 1985, and M.S. and Ph.D. degrees in electrical engineering in 1987 and 1991, respectively, from the University of Notre Dame, Notre Dame, Indiana. He was a Postdoctoral Research Affiliate at the Information Systems Laboratory, Stanford University in 1991. He was the 2001 Cor Wit Visiting Professor at the Delft University of Technology and had held visiting positions at Stanford University, and U.C. Berkeley.

Lang Tong is a Fellow of IEEE. He received the 1993 Outstanding Young Author Award from the IEEE Circuits and Systems Society, the 2004 best paper award (with Min Dong) from IEEE Signal Processing Society, and the 2004 Leonard G. Abraham Prize Paper Award from the IEEE Communications Society (with Parvathinathan Venkitasubramaniam and Srihari Adireddy). He is also a coauthor of seven student paper awards. He received Young Investigator Award from the Office of Naval Research.

Lang Tong's research is in the general area of statistical signal processing, wireless communications and networking, and information theory. He has served as an Associate Editor for the IEEE Transactions on Signal Processing, the IEEE Transactions on Information Theory, and IEEE Signal Processing Letters. He was named as a 2009-2010 Distinguished Lecturer by the IEEE Signal Processing Society. E-mail: ltong@ece.cornell.edu.
\end{IEEEbiography}

\end{document}

%% file: intro_v5.tex
\begin{figure}[htb]
  \centering
  \psfrag{w}[c]{$w_n$}
\psfrag{v}[c]{$e_n$}
\psfrag{u}[c]{$d_n$}
\psfrag{y}[c]{$y_n$}
\psfrag{x}[c]{$x_n$}
\psfrag{u1}[c]{$\Omega_1$}
\psfrag{u4}[c]{$\Omega_2$}
\psfrag{u3}[c]{$\Omega_3$}
\psfrag{L1}[c]{$\Theta_1$}
\psfrag{Lk}[c]{$\Theta_i$}
\psfrag{L2}[c]{$\Theta_K$}
\psfrag{L}[c]{$\Theta$}
\psfrag{t}[c]{$t$}
\psfrag{tk}[c]{$t_k$}
  \includegraphics[width=\hsize]{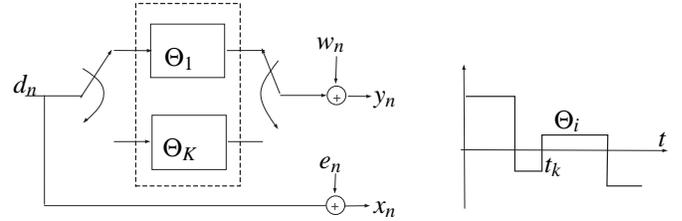}
  \caption{\label{fig1} Left: A jump linear model with a set of linear submodels $\{\Theta_i\}$ and noisy measurements $(x_n,y_n)$.  Right: a trajectory of model evolution with switching points $\{(t_k,\Theta_k)\}$.}
\end{figure}

\PARstart{W}{e} consider the problem of estimating parameters of a deterministic jump linear or piecewise linear model. In a generic form, as illustrated in Fig.~\ref{fig1}, a jump linear model (JLM) is a hybrid system that switches among a set of  multiple input multiple output (MIMO) linear models parameterized by a set of matrices $\Theta=\{\Theta_1,\cdots, \Theta_K\}$.  The underlying mechanism that triggers the switching from one model to another is unknown and is considered exogenous.  The set of switching epochs $\{t_k\}$, however, is part of the unknown parameters to be estimated.

A different model to which the proposed approach is also applicable is the  piecewise linear model (PLM).  For a PLM,  it is the system input that triggers the model switching.    In particular, the input domain of a PLM is partitioned to $K$ subsets $\{\Omega_i\}$, each corresponding to a particular submodel parameterized by some $\Theta_i$.  The switching occurs whenever the system input enters a different partition. A main difference between JLM and PLM is that a PLM associates each system input with a unique linear model, whereas the triggering mechanism of a JLM can produce different system responses for the same input.
In this respect, the JLM is more general and its identification more challenging.

For both models, we consider the following parameter estimation problem:  given a block of noisy observations $(x_n, y_n), n=1, \cdots, N$, estimate system matrices $\{\Theta_i \in \mathbb{R}^{N_y\times N_d}\}$, the model input sequence $d_n \in \mathbb{R}^{N_d}, n=1, \cdots, N$, and the set of switching epochs $\{t_k\}$. Other quantities are either assumed known or treated as nuisance parameters.

As special cases of hybrid systems, a JLM involves parameters of the mixed integer type: the parameter matrices and system input are continuous, the switching epochs  discrete. The lack of continuity in the parameter space makes it difficult to find the globally optimal estimator.  In this paper, we aim at a less ambitious goal of finding a computationally tractable suboptimal technique that, at the minimum, provides exact parameter identification in absence of noise.  And we expect the proposed algorithm performs well at high signal-to-noise ratios (SNRs).

We note that the problem is drastically simplified if we are able to associate each observation $(x_n,y_n)$ with a particular submodel.  Once this is accomplished, we essentially have a classical parameter estimation problem.  Thus at the heart of the problem is one of \textit{unsupervised classification}, knowing that the data are generated in a specific way.

\subsection{Related Work}
JLM and PLM are powerful modeling tools.  They have found applications
in control systems and dynamics \cite{Ferrari_2002,Grieder_2005,Habets_2006,Baric_2008,Richter_2011}, fault detection and diagnosis \cite{Schwaiger_2007,Nayebpanah_2009,Gholami_2011}, system biology \cite{Guha&Biswas2008}, video processing \cite{TuragaEtal2008}, and wireless communications and networking \cite{Doucet2000AC,Biao&Tong2001}.
Thus obtaining model parameters from observation data is of both theoretical and practical significance.

One of the earliest formulation of parameter identification of JLM is by Jiang in \cite{Jiang93CDC}.  If the noise in the model is Gaussian, the optimization problems considered in \cite{Jiang93CDC} leads to the maximum likelihood estimator.  Unfortunately, there is no computationally tractable techniques for the underlying mixed integer optimization.

There is a substantial literature on the identification of piecewise affine models of which PLM is a special case.  See  \cite{Juloski,Paoletti2007} and references therein.
Existing approaches can be classified into three (not necessarily mutually exclusive) categories: clustering techniques,  statistical inference methods, and  algebraic geometry techniques.   The approach proposed in this paper falls in the first category and is most appropriately described as a spectral clustering technique applied to a particular signal subspace.

Clustering based techniques are two-step procedures that first
 associate each observation with a submodel. Once such associations are made,  classical system identification techniques can be applied.  If the data association algorithm has low association error, the performance of such a technique tends to be close to that of the classical system identification technique.  Association errors occur with higher probability at low signal-to-noise ratios (SNR), which causes performance degradation.

When the number of subsystems (input partitions) is known,
various clustering techniques such as the $K$-mean algorithm can be used to group
the observations into sub-regression regions \cite{Ferrari}. Fuzzy clustering combined with competitive learning has also been proposed \cite{Gegundez}. In \cite{Liang2013},  distance information in the regression space is introduced for data association.
   When the number of sub-models is unknown,  a hierarchical clustering technique has been proposed \cite{Hastie,Bemporad2004,Bemporad2005} where the hierarchical clustering can be used to estimate the number of subsystem and system parameters simultaneously. In particular, observation data are first classified into minimal feasible sub-models via a modified greedy algorithm. As the sub-model number determination and parameter estimation are mixed, these methods usually have high computation cost.

The problem of estimating PLM parameters can be casted as a classical statistical inference problem either in point estimation or  Bayesian settings.  Particle filter estimators have been proposed  to obtain the maximum a posteriori (MAP) estimator \cite{Juloski2} or the minimum mean squared error (MMSE) estimator \cite{Doucet2001}. The expectation-maximization  (EM) algorithm has also been applied to estimate parameters of a piecewise affine system  \cite{HayatoNakada} and a Jump Markov linear system \cite{Logothetis1999}.  A generalized expectation maximization (GEM) algorithm is used to separate the data sources and determine the parameters of channels \cite{Routtenberg2009}.

Algebraic geometric techniques is also a two-step procedure where coefficients of a certain polynomials are first estimated and the roots of the polynomial, in the absence of noise, gives the parameters of the piecewise affine system \cite{Vidal,Feng_C}.  These techniques are applicable to systems with a small number of parameters as the number of coefficients to be estimated from product polynomials grows exponentially with the number of the subsystems to be estimated, which means that the minimum number of data samples needed also grows exponentially.  In addition, the computation from polynomial coefficients to roots of the polynomial is sensitive to estimation errors.

The proposed technique is a fusion of two well known ideas in signal processing: one is the exploitation of subspace structure, the other the spectral clustering on a graph.  These two ideas are well studied, but the joint application  to the estimation of JLM/PLM is new.   Relevant subspace methods for system identification can be found in \eg \cite{Viberg:95Automatica,McKelveyAkcayLjung:96IT,TongPerreau:98,Liang&Ding:03TSP,An2005,YuPetropulu:08TSP} and references therein.
%
%
%
%

For spectral clustering techniques, see  a tutorial \cite{Luxberg07SC}.  We adapt in particular the technique of Shi and Malik \cite{ShiMalik2000} for the problem at hand, using a similarity matrix obtained from the subspace decomposition.

\subsection{Summary of results, contributions, and organization}
The estimator proposed in this paper is a spectral clustering technique on a weighted random graph with edge weights obtained from a subspace decomposition, thus we term the algorithm spectral clustering on subspace (SCS).  A key idea is to exploit a subspace structure inherent to switchings of a JLM.  In particular, the jumps in JLM lead to a block diagonal structure of the input subspace, and it is this block diagonal structure that provides a similarity measure among observation data samples. This realization makes it obvious that, in the absence of noise, the proposed technique gives the exact parameter identification.

While the proposed SCS algorithm appears to be the only technique that provides exact parameter identification for JML models, it has both advantages and weaknesses when it is compared with existing algorithms.  Comparing with its close relatives in the clustering-based methods,  the proposed spectral clustering technique is applicable to  input domains with arbitrarily  shaped (non-convex and possibly non-connected) partitions.  Standard clustering techniques (such as the $K$-mean algorithm) applied directly in the observation space partition the input domain into convex subregions.  In addition, standard clustering techniques may have classification errors even in the absence of observation noise. When compared with algebraic geometry techniques and statistical inference methods in the literature, the SCS algorithm is based on the more numerically stable SVD and deals more easily with MIMO systems.  In simulations, SCS achieves the ``clairvoyant'' Cram\'{e}r-Rao bound at high SNR, to which both the clustering and algebraic geometry methods have a noticeable gap even at very high SNRs.

There are  prices paid for the advantages of the proposed approach.    A rank condition assumed in  A4 of Section~\ref{Sec:2} requires, at the minimum, that total number of input and output sensors $(N_d+N_y)$ has to be greater and equal to $K\times N_d$, the product of the number of subsystems and the number system inputs.  Not required by some of the existing techniques, this assumption is essential in identifying subspaces associated with each sub-model; it is  a physical limitation in its applications.  A second weakness is that the subspace method used here has a SNR threshold below which the performance degrades.  The breakdown threshold depends on the parameters and specific applications. We do not have a characterization of the breakdown threshold.  We note that this weakness is not unexpected as many subspace techniques have this unfortunate characteristic.

The remainder of the paper is organized as follows: Section II presents the system model and the problem formulation. Here we present the set of assumptions made in this paper and discuss their implications.
Section III presents the key result on a block diagonal structure of the observation subspace.  We present first the idea using the scaler example and give the full characterization in Theorem~\ref{thm1}.
Section IV proposes graph spectral clustering technique that partition data samples into groups, each is associated with a particular subsystem model.
 Section V presents  numerical  studies and some comparisons with existed methods. The last section VI gives the conclusions.

Throughout the paper, the following notations are adopted:
\begin{center}
\begin{tabular}{|ll|}
\hline
    $N$                                    &the number of observations\\
    $K$                                    &the number of subsystems\\
    $d_n$                               &input of system\\
    $d_n^{(i)}$                         &input of $i$th submodel\\
    $y_n, Y_n$                          &output observations\\
    $x_n, X_n$                          &input observations\\
    $z_n, Z_n$                          &extended vector of observations\\
    $e_n$                               &input noise\\
    $w_n$                               &output noise\\
    $\theta_i, \Theta_i$                &parameters of $i$th submodel\\
    $\Omega_i$                          &domain of $i$th submodel\\
    $\ell_i$                            &label set of $i$th submodel\\
    $A$                                 &extended parameters matrix\\
    $V$                                 &subspace of right-singular matrix of $Z$\\
    $W$                                 &adjacency matrix with $W=|VV^{\T}|$\\
    $\Pmsc$                             &domain partitions\\
    $P$                                 &permutation matrix\\
    $\otimes$                           &Kronecker product\\
 \hline
\end{tabular}
\end{center}

%% file: ProblemFormulation_v4.tex
In this section, we describe a JLM  as illustrated in Fig.~1, present main assumptions made in this paper, and discuss implications of our assumptions.

A jump linear model index by deterministic parameters is denoted by ${\cal M}(\Tmsc,\Theta,\bar{D})$, where $\Tmsc=\{\Tmsc_i, i=1, \cdots, K\}$ is a partition of the overall time horizon
$\{1,\cdots, N\}$ during which the observations are made where $\Tmsc_i$ is the set of time indices that the JLM operates as the subsystem $i$ with parameter matrix $\Theta_i$. Matrix $\Theta=[\Theta_1,\cdots, \Theta_K]\in \mathbb{R}^{N_y \times KN_d}$ contains all submodel system matrices, and input matrix $\bar{D}=[d_1,\cdots, d_N]\in \mathbb{R}^{N_d\times N}$ includes the deterministic input vector.

The observation model of ${\cal M}(\Tmsc,\Theta,\bar{D})$ is defined by
\begin{eqnarray}
    x_n &=& d_n+e_n,  \label{eq:model1}\\
    y_n &=& \left\{\begin{array}{ll}
    \Theta_1 d_n + w_n, &  n \in \Tmsc_1\\
    \cdots & \cdots \\
        \Theta_K d_n + w_n, &  n \in \Tmsc_K\\
    \end{array} \right..  \label{eq:model2}
\end{eqnarray}
Here all input and output variables are vectors, and parameters $\Theta_i$ are matrices of compatible dimensions.

We can write the above in a more compact form by defining
\begin{eqnarray}
z_n &\triangleq & \left[\begin{array}{c}
                                      x_n \\
                                      y_n
                                    \end{array}
  \right],~~~Z\triangleq\left[z_1, \ldots, z_N\right]\\
A(\Theta)&\triangleq & \left[\begin{array}{ccc}
                                                 I & \ldots & I \\
                                                 \Theta_1 & \ldots & \Theta_N
                                               \end{array}
  \right],\\
  D & \triangleq & \mbox{diag}\left(D_1, \ldots, D_K\right),
\end{eqnarray}
where $D_i=[d_1^{(i)},\cdots, d_{N_i}^{(i)}]$ is the matrix of ordered (according to the time of arrival) input vectors of the $i$th subsystem, \ie $d_{k}^{(i)}$ is the  $k$th input
associated with the $i$th submodel.  We then have
\begin{equation}
Z=A(\Theta)DP + E
\end{equation}
where $P$ is a permutation matrix that re-orders the columns of $D$ to match the arrival sequence in $Z$ and $E$ the matrix containing measurement noise samples.

Finally, we note that, while the model given in (\ref{eq:model1})-(\ref{eq:model2}) appears to be for a static system, by allowing the input vector $d_n$ to include past observations, the model  includes MIMO systems with finite impulse responses.  For such models, matrices $\Theta_i$ have a block Toeplitz structure that can be exploited, which we do not consider in this paper.

Main assumptions made in developing our approach and analysis are as follows:
\begin{itemize}
\item[A1:] We assume the observation noise $(e_n, w_n)$ is i.i.d. Gaussian with  zero mean and covariance matrix $\mbox{diag}(\sigma_x^2 I, \sigma_y^2 I)$.  This is a mild assumption.  In fact, our algorithm does not make use of the Gaussian assumption; A1 is made for the purpose of deriving a Cram\'{e}r-Rao like lower bound against which the proposed algorithm and other benchmarks can be compared.

\item[A2:] We assume that
the number $K$ of submodels  and the dimensions of submodels are fixed and known.   This is an assumption that most existing techniques make and is quite restrictive. Identifying the number of submodels and their respective dimensions are challenging although there are many practical techniques including some based on eigenvalue decompositions.

\item[A3:]  We assume that the input matrix $D$ has full row rank.   We further assume that each sub-block $D_i$ cannot be row-permutated into a diagonal form.

     The full rank condition on $D$ requires that the each subsystem is persistently excited---a necessary condition for identifiability assuming that we can associate observation data samples $\{z_n\}$ perfectly with the submodels $\{\Theta_i\}$ that generates them.

     The condition on individual input $D_i$ not permutable to diagonal form is also necessary for identifiability.   If $D_1=\mbox{diag}(D_{11}, D_{12})$, then we can break up $A(\Theta_1)$ accordingly into $A(\Theta_{11})$ and $A(\Theta_{12})$ and group $(\Theta_{12},D_{12})$ with the second submodel $(\Theta_2,D_2)$.  The system then becomes unidentifiable.

\item[A4:]  We assume that matrix $A(\Theta)$ has full column rank.  The significance of this assumption will become clear in Section~\ref{Sec:3}.  It is evident that this  assumption limits the applicability of the proposed algorithm.
    In particular, it implies that, for the proposed technique to be applicable, the number of input output sensors must scale linearly with the number $K$ of subsystems.
\end{itemize}

%% file: subspace_v4.tex
We present in this section a subspace structure in the observation that shows the decomposition for data associated with different sub-models followed by algorithms that label input-output data. Once the labels are obtained, the problem of estimating system parameters becomes standard.

\subsection{Subspace decomposition: a sketch of ideas}
The subspace structure can be easily described using the following simple noiseless {\em scaler} bi-model JLM:
\begin{eqnarray}
      x_n &=& d_n, \label{eq:simple_a}\\
    y_n &=& \left\{\begin{array}{ll}
    \theta_1 d_n &  n \in \Tmsc_1\\
    \theta_2 d_n & n \in \Tmsc_2\\
    \end{array} \right.  \label{eq:simple_b},
\end{eqnarray}
where $\Tmsc_1$ and $\Tmsc_2$ are arbitrary disjoint subsets of $\{1, \cdots, N\}$.

Consider, for the moment, the special case when the system input arrives sequentially, the first $N_1$ samples from sub-model 1 and the next $N_2=N-N_1$ from sub-model 2.  In this special case, we have
\begin{equation} \label{eq:Z=AU}
Z=AD,
\end{equation}
where
\begin{eqnarray}
A &\defeq & \left[\begin{array}{cc}
  1 & 1\\
    \theta_1&\theta_2\\
  \end{array}\right], \\
D & \defeq &  \left[\begin{array}{cccccc}
    d_1^{(1)}&\ldots&d_{N_1}^{(1)}&0&\ldots&0\\
    0&\ldots&0&d_{N_1+1}^{(2)}&\ldots& d_N^{(2)}
  \end{array}\right]\nonumber\\
&=&  \mbox{diag}(D_1, D_2)  \label{eq:diagU}
\end{eqnarray}
where vector $D_1=[d_1^{(1)},\cdots, d_{N_1}^{(1)}]$ from the first sub-model
 with parameter $\theta_1$ and the next $N_2$ samples form vector  $D_2$  from the second sub-model.

Let $Z$ have the singular value decomposition of the form
\begin{equation}
  Z =Q\Sigma V^{\T},
  \end{equation}
Note that $\Sigma>0$ is a diagonal matrix with reduced dimension $2\times 2$, $Q$ is an orthogonal matrix, and
 $V^{\T}$ is a $2\times N$ matrix of the same dimension of input matrix $D$.  Indeed, when $A$ is nonsingular,
$V^{\T}$ spans the same row space as $D$.  In other words, we have
\begin{equation} \label{eq:VT=TU}
V^{\T}=TD,~~T\defeq \Sigma^{\text{\tiny -1}}Q^{\T}A.
\end{equation}
Because $V^{\T}V=TDD^{\T}T^{\T}=I$ and $T$ is nonsingular,
\begin{equation} \label{eq:TT}
T^{\T}T=\left(DD^{\T}\right)^{\text{\tiny -1}}.
\end{equation}
Thus, we have
\begin{align}\label{eq:VVT}
VV^{\T} = D^{\T}T^{\T}TD&=D^{\T}\left(DD^{\T}\right)^{\text{\tiny -1}}D\notag\\
        &= \left[\begin{array}{cc}
                  \frac{1}{||D_1||^2}  D^{\T}_1D_1 & 0 \\
                  0 & \frac{1}{||D_2||^2} D_2^{\T}D_2\\
                \end{array}
                \right].
\end{align}
Therefore, matrix  $VV^{\T}$ is block diagonal with the first block made of data from only the first sub-model  and second block from only the second sub-model. The key observation is that matrix $VV^{\T}$ gives the partition of input sequence which is exploited in the proposed algorithm.

If the system input sequence is not arranged in the block sequential fashion as in (\ref{eq:diagU}), matrix $VV^{\T}$ is no longer block diagonal; instead it will be a block diagonal matrix under a certain similar transform defined by a permutation matrix $P$.  However, the permutation induced transform can be easily reversed to recover the block diagonal structure.

\subsection{Subspace decomposition}
We now formally state the subspace decomposition by removing the sequential arrival restriction on the input sequence.
The following theorem captures the subspace structure in the row space of data matrix $Z$.
\begin{theorem}\label{thm1}
Consider the noiseless system model given by
\begin{equation}
  \label{Z=AUP}
  Z=A(\Theta)DP,
\end{equation}
where $P$ is an unknown permutation matrix that permutes the ordered input sequence to the actual sequence of arrivals.    Let $Z$ have the singular value decomposition of the form
  \begin{equation}
    \label{SVD}
    Z=Q\Sigma V^{\T},
  \end{equation}
  where $QQ^T=I$ and $V^{\T}$ of the same size as $D$ with orthogonal rows. Under assumptions (A3-A4),
  \begin{equation}
  \label{Lambda}
    VV^{\T}=P^{\T}\mbox{diag}\left(\Lambda_1,\ldots, \Lambda_N\right)P,
  \end{equation}
  where $\Lambda_i=D^{\T}_i \left(D_iD_i^{\T}\right)^{-1}D_i$.
\end{theorem}

\vspace{0.5em}
\begin{IEEEproof}
See Appendix.
\end{IEEEproof}

\vspace{0.5em}
\begin{remark}
If the system input sequence arriving in consecutive from the same sub-model, we have $P=I$.  In this case, matrix $VV^{\T}$ is block diagonal. In general, matrix $P$ scrambles $\mbox{diag}(\Lambda_1,\cdots, \Lambda_N)$.  For noiseless measurement, the matrix can be easily de-scrambled by permuting $VV^{\T}$ back to diagonal form.  See \cite{Aykanat&Pinar&Catalyurek:04SIAM} for an efficient de-permutation algorithm.  In the next section, we present a spectral clustering algorithm that recovers the structure in (\ref{Lambda}) in the presence of measurement noise.
\end{remark}

\begin{remark}  While the above theorem reveals the structure that can be exploited, it is not yet sufficient for correctly identifying the JLM. In particular, since input sequence in $D_i$ may contain zero entries, and it is not clear that $\Lambda_i$ does not contain additional block diagonal entries.  If that were the case, $VV^{\T}$ would not have been enough for identifying uniquely the input subspaces with submodels.  This issue is addressed later in Section~\ref{property}.
\end{remark}

%% file: SpectralCluster_v4.tex
In this section, we recast the problem of data association as an inference problem involving a random graph.    The key idea is to represent data by a data association (or similarity) graph with data samples  as its vertices and its edges representing whether observations are generated from the same  subsystem.  Such a representation partitions the graph into groups with each group associated with a particular subsystem.  The problem of data association then becomes estimating the graph structure.  To this end, we adapt a spectral clustering technique based on eigenvectors of normalized graph Laplacian.

\subsection{A Random Graph Representation}
We define a data associate graph $\Gmsc=(\Vmsc,\Emsc)$ with random edge weights where
$\Vmsc=\{1,2,\cdots,N\}$ is the set of vertices with vertex $i$ corresponding to observation sample $(x_i,y_i)$ and $\Emsc=\{(i,j), \forall i,j \in \Vmsc\}$ the set of undirected edges.  The edge weight matrix $W$ is random and  is defined by the SVD of the observation $Z$ in (\ref{SVD}).  In particular, edge weight matrix is given by
\[
W \defeq |VV^{\T}|,
\]
where the operation $|A|$ takes the absolute value on each entry of matrix $A$.  Two vertices are connected if and only if there is an edge with positive edge weight.

In light of (\ref{Lambda}), in the absence of noise, the weight of the graph is given by
\[
W = P^{\T}\mbox{diag}(|\Lambda_1|,\cdots, |\Lambda_k|)P.
\]
If we assume that entries of $\Lambda_i$ are non-zero, which may not be true in practice,  the graph representing the ground truth has exactly $K$ components, homomorphic to the set of $K$ submodels of JLM.

With measurement noise, entries of $W$ are almost surely non-zero.  Nonetheless, the $(i,j)$th entry $w_{ij}$ of $W$ can be viewed as the strength of the evidence that data observed at tie $i$ and $j$ are generated by the same subsystem.

The problem of determining data association can then be casted as one of making inference on the connectivity structure of the graph given the realization of random weight matrix $W$, for which we propose a spectral clustering technique.

\subsection{Structural inference via spectral clustering}
Given the data observation graph $\Gmsc$ and its weight matrix $W$, the problem of data association is to partition $\Gmsc$ into $K$ connected components.  Each component contains only data from one and only one subsystem.

There are a number of spectral clustering techniques that are applicable for the problem at hand.  Here we illustrate the main idea of one such approach.  The interested reader may find a detailed exposition from \cite{Luxberg07SC} and references therein.

Consider the simple noiseless case when $K=2$.  Without loss of generality, we ignore the permutation matrix in (\ref{Lambda}), \ie  $W=\mbox{diag}(|\Lambda_1|,|\Lambda_2|)$.

A normalized graph Laplacian can be defined as
\[
\bar{L}=I-R^{-1}W,~~~R=\mbox{diag}(\sum_j w_{1j},\cdots, \sum_j w_{Nj}).
\]
It can be easily verified that  $\bar{L}$ has exactly $K=2$ eigenvalues equal to zero, and the eigenspace associated with the zero eigenvalue is spanned by a set of $K=2$ indicator vectors $\{\delta_1, \delta_2\}$ where  $\delta_i$ has ones at entries corresponding to data from subsystem $i$ and zero elsewhere.  In other words, eigenvector $\delta_i$  associated with the zero eigenvalue identifies exactly the data samples associated with subsystem $i$.

The above idea generalizes to the $K$ subsystem case. In the absence of noise, there are $K$ eigenvectors associated with the zero eigenvalue, each eigenvector whose non-zero entries  indicates observations associated with the same subsystem.

%% file: SCA_v2.tex
In this section, we present an implementation of the SCS algorithm.  The algorithm includes three parts: (i) compute the weight matrix of a random graph from the SVD of the signal space; (ii) apply a spectral clustering algorithm to associate each sample with a subsystem.  The association provides estimates of the jumping epochs; (iii) estimate the system matrices and system input by the total least squares algorithm or via the maximum likelihood estimation.  The specifics of the implementation is shown in Algorithm~\ref{alg1}.

\begin{algorithm}
  \caption{\label{alg1} The Spectral Clustering on Subspace (SCS)}
  \begin{algorithmic}
  \REQUIRE \STATE Measurements $X=\{x_n\}_{n=1}^{\N},Y=\{y_n\}_{n=1}^{\N}$
  \ENSURE \STATE Parameter estimations $\{\hat{\Theta}_i\}_{i=1}^{\K}$, label set $\{\ell_i\}_{i=1}^{\K}$
    \STATE (1) compute the SVD of measurement matrix $Z=[X;Y]$
    \STATE    \qquad\qquad$Z=Q\Sigma V^{\T}$
    \STATE (2) calculate adjacency matrix $W$ and Laplacian matrix $L$:
    \STATE     \qquad\qquad$W=|VV^{\T}|$, $L\triangleq R-W$
    \STATE     \qquad\qquad$R\triangleq diag(\sum_{j=1}^Nw_{1,j}, \ldots, \sum_{j=1}^Nw_{N,j})$
    \STATE (3) calculate the first $K$ generalized eigenvectors of $(L,R)$
    \STATE     \qquad\qquad$L\delta_i=\lambda_iR\delta_i, i=1,\ldots, K$
    \STATE (4) let $\Delta\triangleq\left[\delta_1,\ldots,\delta_K\right]$, cluster the rows of $\Delta$ into $K$ groups with $K$-means:
               \STATE\qquad\qquad $[cent, ind]=kmeans(\Delta,K)$
               \STATE\qquad\qquad $\ell_i=find(ind==i), i=1,\ldots, K$
    \STATE (5) estimate the parameters via total least squares
    \STATE     \qquad\qquad$\hat{\Theta}_i=tls(X(\ell_i),Y(\ell_i)), i=1,\ldots, K$
  \RETURN $\{\hat{\Theta}_i\}_{i=1}^{\K}$, $\{\ell_i\}_{i=1}^{\K}$
\end{algorithmic}
\end{algorithm}

\begin{remark}
In the absence of noise, the $K$ eigenvectors are indicator functions that directly associate each observation to a subsystem.  With noise, these eigenvectors no longer contain entries exactly equal to one or zero.  In our implementation, each row of $K$ eigenvectors is taken as a joint
 indicator for subsystems, then all rows are grouped into $K$ clusters by $K$-mean.
\end{remark}

%% file: Performance_v4.tex
We address performance issues in this section by considering two cases.  First, we establish the identifiability for the noiseless case, which shows that the proposed algorithm will provide exact identification of JLM parameters.
Second, in the presence noise, we derive a clairvoyant Cram\'{e}r-Rao bound (C-CRB) by assuming that we know $\Tmsc$ perfectly.  

We should point out that an efficient estimator does not exist in this case. The use of the C-CRB is reasonable only at high SNR because the proposed algorithm tends to obtain perfect $\Tmsc$ estimate and is approximately unbiased.

\subsection{A Sufficient Condition on Identifiability\label{property}}
In the absence of noise, we consider whether parameters of a JML model ${\cal M}(\Tmsc,\Theta,\bar{D})$ can be uniquely determined from the observation matrix $Z$ under assumptions A3-A4.  

The following  corollary of Theorem~\ref{thm1} specifies a sufficient identifiability condition on the system input achieved by the SCS algorithm. Essentially, the condition requires that the graph associated with the input sequence of each subsystem forms a single component.  The proof is immediate hence omitted.

\begin{corollary}\label{corl}
Assume that the input matrix $D_i$ of submodel $i$ does not contain zero columns for all $i$.  A JLM $M(\Tmsc,\Theta,d)$ is uniquely identified (up to a permutation of submodels) by the SCS algorithm if, for all $i$, the smallest eigenvalue ($0$) of the Laplacian of the input graph of the $i$th submodel  with the weight matrix  $W_i\defeq |D_i^{\T}(D_iD_i^{\T})^{-1}D_i|$  has multiplicity $1$.
\end{corollary}

The above Corollary states a one-one correspondence between the components of the observation graph defined by the subspace spanned by $V$ and those of the system input graph defined by that spanned by $D$.
Note that the assumption that input matrix $D_i$ does not contain zero columns is made without loss of generality, given A3.  In particular, the problem of identifiability is not affected by input vectors that are zero.

It is instructive to consider some special cases of input sequences that make the system identifiable or not identifiable by the proposed algorithm. First, for single-input multiple output (SIMO) JLMs, in the absence of noise, the SCS algorithm always correctly identifies the true JLM parameters. The same is true for MIMO JLMs whose input sequence $d_n \in R^M$ are chosen randomly with some continuous distribution on a subset of $R^M$ with positive Lebesgue measure.

It is also not difficult to construct input sequences not satisfying the condition in Corollary~\ref{corl} that the SCS algorithm is expected to fail even in the absence of noise.  One such  case is given by $D_1=[p, \cdots, p, q, \cdots, q]$ where $p^{\T}q=0$. In this case, the weight matrix of the graph associated with the input of subsystem 1 has two components.  The graph associated  with observation subspace then has $K+1$ components with an extra component that can be grouped with any of the one of the $K$ components, causing possible indeterminacy by the SCS algorithm.

\subsection{Clairvoyant Cram\'{e}-Rao Bound\label{C-CRB}}
We derive in this section a ``clairvoyant'' CRB (C-CRB) by assuming that we know $\Tmsc$--the partition of the time horizon with respect to submodels.  In this case, only samples in $\Tmsc_i$ are observations associated with submodel $i$, and the estimation of parameters $(\Theta_i, D_i)$ depends only $Z_i=[z_{n_1},\cdots, z_{n_{N_i}}],  n_k \in \Tmsc_i$.

Without loss of generality, and for notation brevity, we drop the submodel index $i$ and consider the following model that applies to any submodel:
\begin{eqnarray}
x_t &=& d_t + e_t,~~~t=1,\cdots, N\\
y_t &=& \Theta d_t + w_t = {\cal H}(d_t)\theta + w_t
\end{eqnarray}
where $\Theta \in \mathbb{R}^{N_y\times N_d}$ and ${\cal H}(d_t). \triangleq I \otimes d_t^{\T} \in \mathbb{R}^{N_y \times N_yN_d}$.  Denote $\bar{D}=[d_1,\cdots, d_N]$.

  Under A1, we have
independent noise sequences $e_t \stackrel{\mbox{\tiny i.i.d.}}{\sim} {\cal N}(0, \sigma_e^2 I)$ and $w_t \stackrel{\mbox{\tiny i.i.d.}}{\sim} {\cal N}(0, \sigma_w^2 I)$.  The unknown deterministic parameters are $\overrightarrow{\theta} \triangleq \mbox{vec}(\Theta^{\T})=[\theta_1^{\T},\cdots,\theta_{N_y}^{\T}]^{\T}$ and  $\overrightarrow{d} \triangleq \mbox{vec}(\bar{D})$.

The Fisher information matrix for $F(\overrightarrow{\theta},\overrightarrow{d})$ can be easily derived and shown to have the following form
\[
F(\overrightarrow{\theta},\overrightarrow{d}) =
 \left[\begin{array}{cccc}
F_{\theta} & F_{\theta,d_1} & \cdots & F_{\theta,d_N}\\
F_{\theta,d_1}^{\T} & F_{d_1} &  & \\
\vdots  & & \ddots & \vdots \\
F_{\theta,d_N}^{\T} & & \cdots & F_{d_N}\\
\end{array}
\right]
\]
where
\begin{eqnarray}
F_{\theta} &=& \frac{1}{\sigma_w^2}\sum_i {\cal H}^{\T}(d_i){\cal H}(d_i)= \frac{1}{\sigma_w^2} I \otimes (DD^{\T})\\
F_{d_i} &=& \frac{1}{\sigma_e^2}I + \frac{1}{\sigma_w^2} \Theta^{\T}\Theta,\\
F_{\theta,d_i}^{\T} &=& \frac{1}{\sigma_w^2} \Theta^{\T}{\cal H}(d_i) =\frac{1}{\sigma_w^2}[\theta_1 d_i^{\T},\cdots, \theta_{N_y}d_i^{\T}].
\end{eqnarray}
The C-CRB for $\theta$ and $d_i$ are then given by
{\small \begin{eqnarray}
\mbox{Cov}(\theta)&\ge&\sigma_w^2 \left(I \otimes (DD^{\T})
-\sigma_e^2 (\sum_t {\cal H}^{\T}(d_t)\Sigma_\Theta{\cal H}(d_t))\right)^{-1}\\
\mbox{Cov}(d_i) &\ge& \sigma_e^2\left(I+\frac{\sigma_e^2}{\sigma_w^2}\Theta^{\T}\Theta -\frac{\sigma_e^2}{\sigma_w^4}\Theta^{\T}{\cal H}(d_i)C_i{\cal H}^{\T}(d_i)\Theta \right)^{-1},
\end{eqnarray}}
where $\Sigma_\Theta\triangleq \Theta(\sigma_w^2I+\sigma_e^2\Theta^{\T}\Theta)^{-1}\Theta^{\T}$ and
$C_i$ the C-CRB of $\theta$ without sample $z_i$.

Note that when $\sigma_e^2\rightarrow 0$, the expressions above converge to  the conventional expressions for linear models.   The case when $\sigma_w^2 \rightarrow 0$ is less obvious.  When $\sigma_w^2=0$, we have $y_n=\Theta d_n$, which impose a constraint on parameter $\Theta$ and $d_n$.  To compute the CRB for this case, we need to pose this as a constrained optimization problem. See \cite{StoicaNg:98SPL}.

%% file: Simu_v6.tex
In this section, we present numerical studies and compare the proposed algorithms with two benchmark techniques.  One benchmark technique is the $K$-mean clustering based algorithm in \cite{Ferrari} which assumed that local data are likely belong to the same subsystem. Sample data are transformed into feature space by defined a local data set (LD), and the $K$-mean algorithm is adopted to cluster the feather vectors. The $K$-mean estimator is an iterative scheme.  To avoid local optimal solution, the implementation repeats the clustering process multiple times, each with a new set of initial centroids. 

The second benchmark is an algebraic geometric technique based on the hybrid decomposition constraints (HDC) in \cite{Vidal}.  Algebraic geometric technique embeds the parameters of subsystems into a polynomial which is estimated by the least squares algorithm, then recover them through a polynomial differentiation algorithm (PDA). The system parameters are then computed via a polynomial differential algorithm (PDA).

We use the standard mean squared error (MSE) as the performance metric in comparing different estimators.  Here the statistical average is taken with respect to random measurement noise in the system input and output.
In Monte Carlo simulations, sample MSE is used as an estimate of the actual MSE. The system input sequence $d_n$ is considered deterministic and is fixed in the Monte Carlo simulations.

We are interested in the performance of different estimators at different levels of signal-to-noise ratio (SNR). In particular, we define SNR (in dB) as
\[
\mbox{SNR}=10 \log  \frac{\sum_{n=1}^N  \left(\sum_{k=1}^K ||A(\Theta_k)d_n||^2 \times \pi_{n,k}+ ||d_n||^2\right)}{N(N_x\sigma_e^2+N_y\sigma_w^2)}
\]
where $\pi_{n,k}\in \{0,1\}$ is an indicator for submodel $k$ which takes $1$ only if the $d_n$ is applied to subsystem $k$ and $0$ otherwise. In the denominator, $N_x$ is the dimension of vector $x_n$ and $N_y$ the dimension of $y_n$.  Intuitively, the numerator is the total signal energy in the input and out sequence and the denominator the total noise energy.

We also compare the performance of various estimators using ``clairvoyant'' Maximum Likelihood  (C-ML) and a  ``clairvoyant'' Cram\'{e}r-Rao bound (C-CRB) where we assume that labels of observation data are known.  The calculation of ``clairvoyant'' Cram\'{e}r-Rao bound is attached in \ref{C-CRB}.

\subsection{SISO JLM with two subsystems\label{case1}}
The first numerical case was  a bi-model jump linear systems with scalar parameters:
\begin{equation}
\begin{split}
  y_n&=\left\{\begin{array}{ll}
    0.7d_n+w_n&n \in \Tmsc_1\\
    0.8d_n+w_n&n \in \Tmsc_2
  \end{array}\right.\\
  x_n&=d_n+e_n,
\end{split}
\end{equation}
where $w_n\sim N(0,\sigma_1^2)$, $e_n\sim N(0,\sigma_2^2)$ are i.i.d. random sequences. A similar example was used in \cite{Ferrari,Vidal} except that we have made the scenario more challenging by making the parameters of the two subsystems relatively close. In the simulation, the input data were generated by an uniform distribution on $[-1,1]$ and kept fixed in Monte Carlo runs.

\begin{figure*}[thpb]
  \centering
  \includegraphics[width=0.48\hsize]{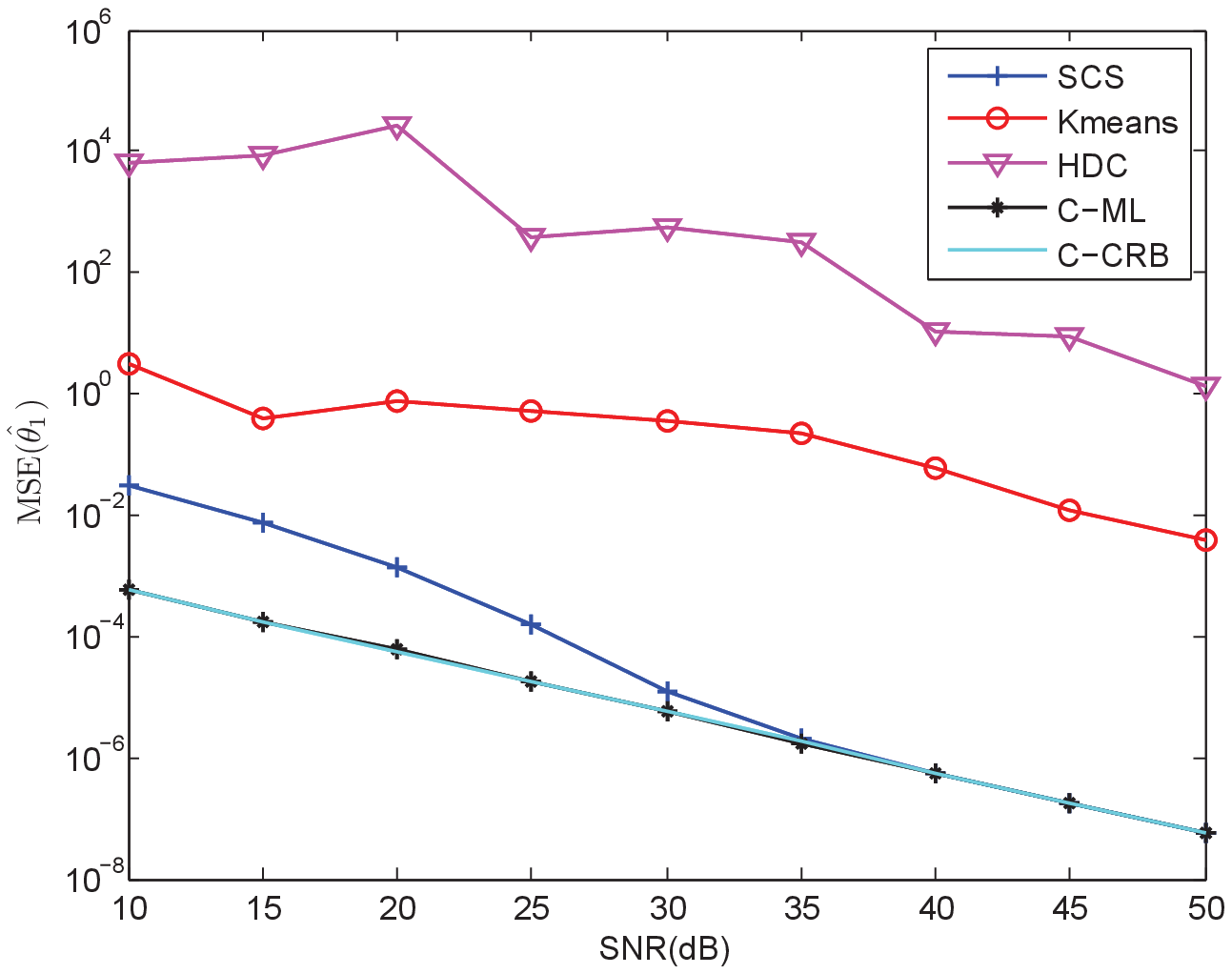}
  \includegraphics[width=0.48\hsize]{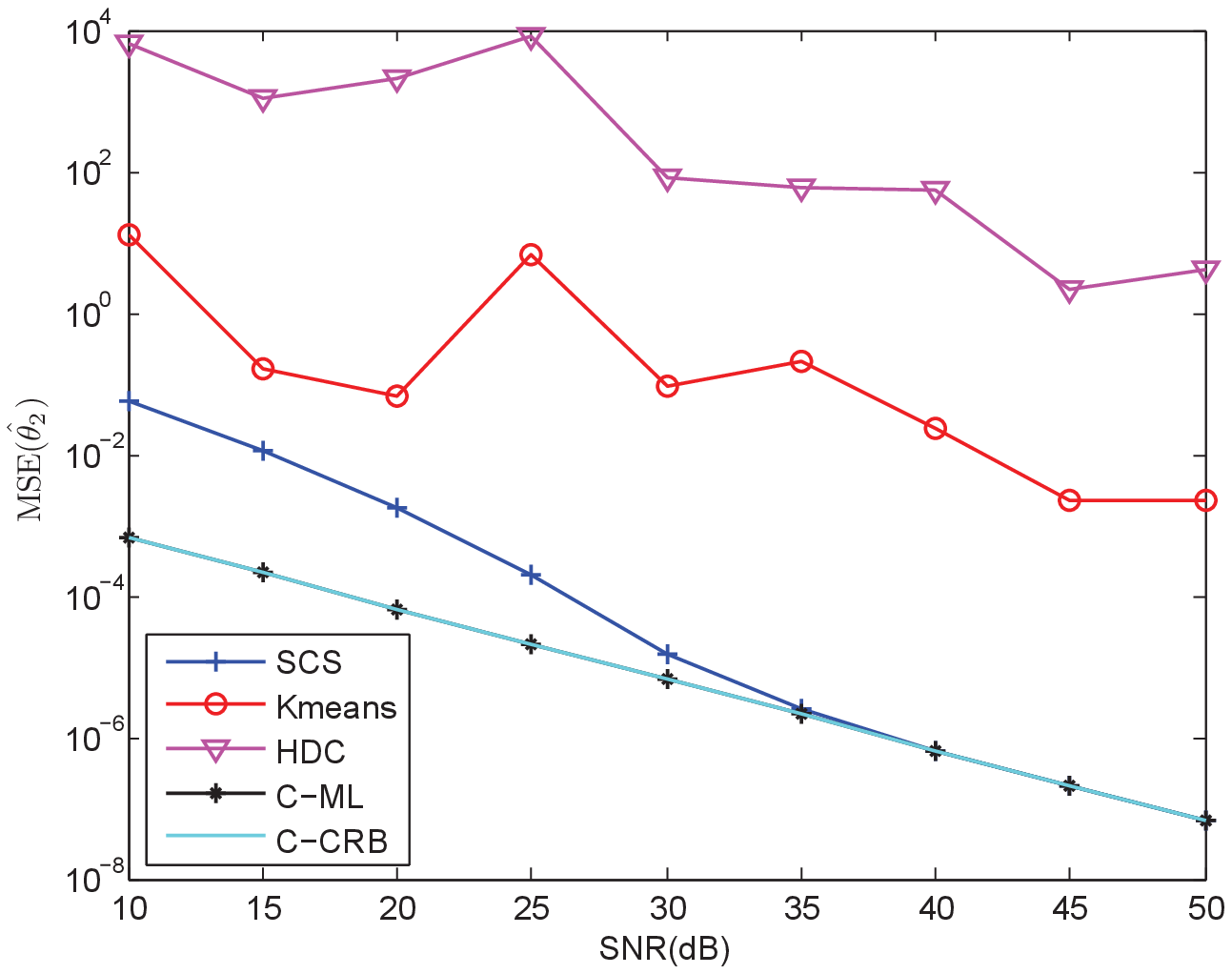}
  \caption{\label{fig_1D_mse} Mean Squared Error (MSE) vs. SNR plots in Example 1. Sample amount is $N_1=200$, $N_2=200$, Monte Carlo runs is $10^4$. SCS: the spectral clustering on subspace algorithm; Kmeans: the $K$-mean based method in \cite{Ferrari}, the tuned coefficient of local data set was set as $c=4$; HDC: algebraic method based on \textit{hybrid decoupled constraints} \cite{Vidal}.
  C-ML: the maximum likelihood solution with the labels of observation known; C-CRB: the ``clairvoyant'' Cram\'{e}r-Rao bound.}
\end{figure*}

\begin{figure*}[thpb]
  \centering
  \includegraphics[width=0.48\hsize]{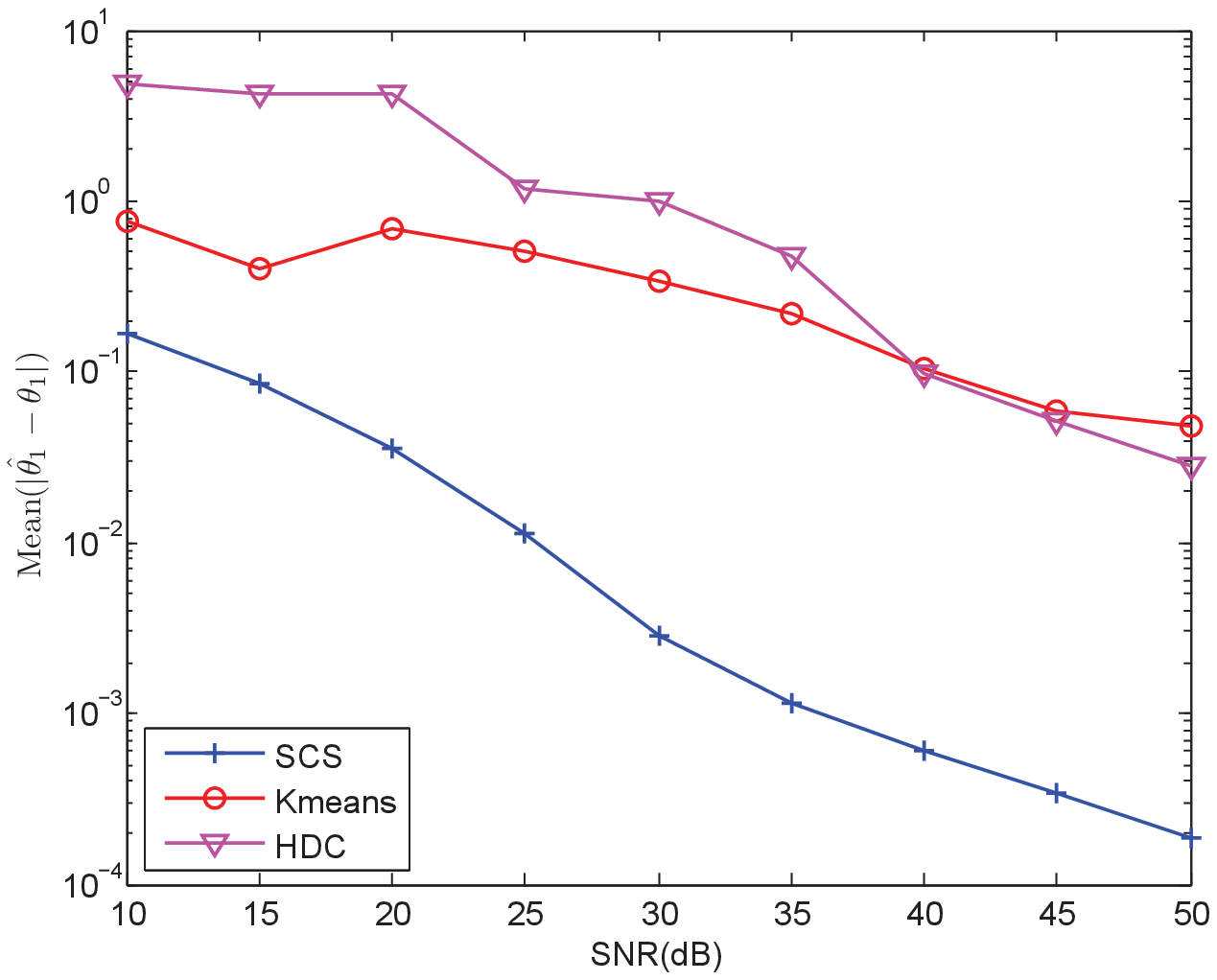}
  \includegraphics[width=0.48\hsize]{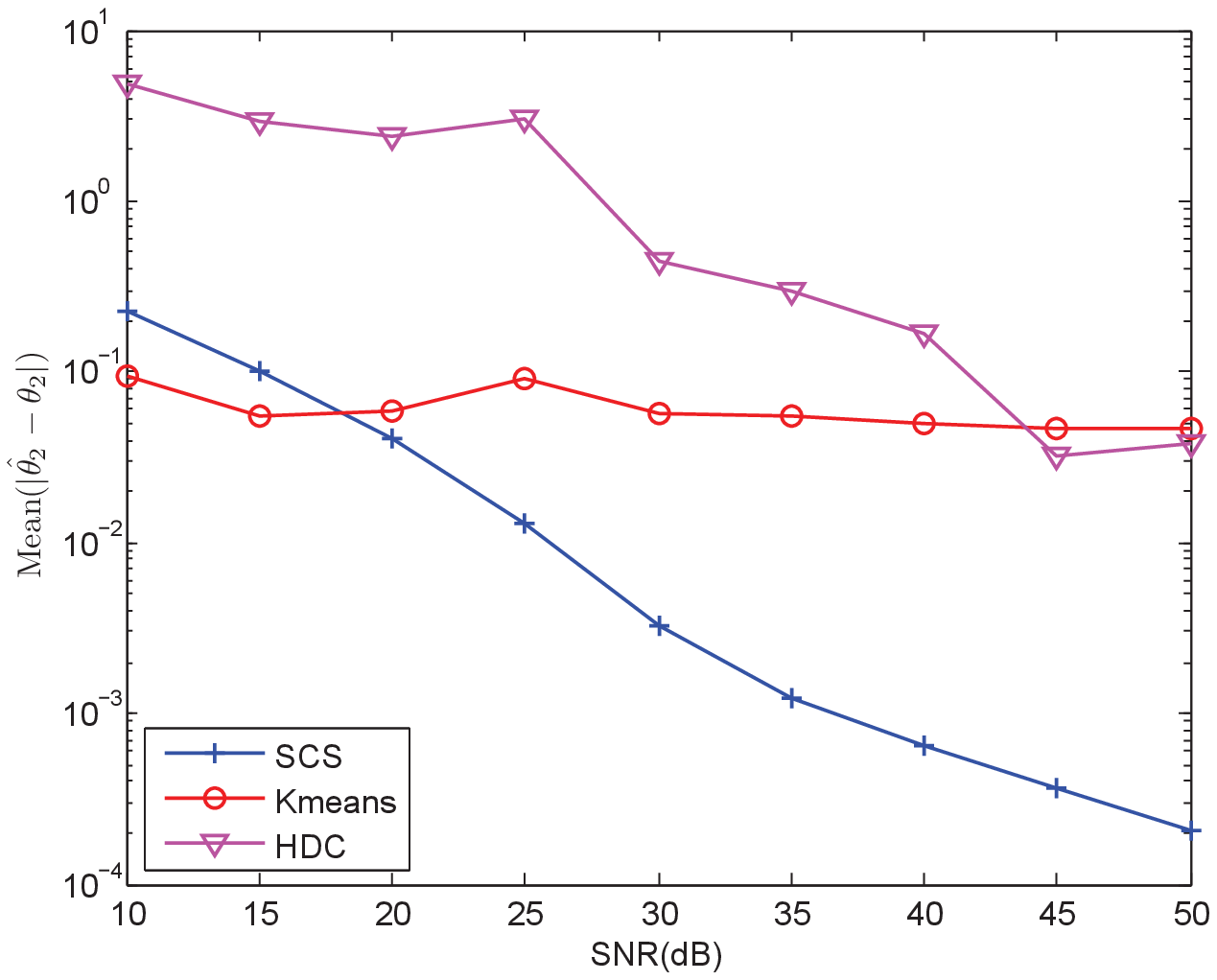}
  \caption{\label{fig_1D_mean} Means of Estimations $\theta_1$ and $\theta_2$ in Example 1.}
\end{figure*}

Fig.~\ref{fig_1D_mse} shows the MSE plots of the SCS algorithm (SCSA) against the benchmarks and (clairvoyant) CRB.
We observed that the proposed method achieved the ``clairvoyant'' CRB at SNR greater than $35$dB whereas the other methods have a gap to the lower bound at very high SNR. As expected, the subspace method had an SNR threshold above which its performance was very competitive, in this scenario the threshold is $35$dB. The key assumption for $K$-mean based method in \cite{Ferrari} is that local data tend to belong the same subsystem. In this simulation, both subsystems were defined on the identical domain. $K$-mean based method did not perform well in this scenario. The algebraic method, referred to as the hybrid decomposition constraints (HDC) algorithm, is applicable to this example and is expected to have perfect identification in the absence of noise.  However, the HDC algorithm appeared to be sensitive to the presence of the noise and did not perform well for SNR levels below 50dB.  (The HDC algorithm performed well when SNR$>50$dB). Fig.~\ref{fig_1D_mean} shows the biases of the three tested algorithms. Similar trends as those in MSE were observed.


\begin{figure}[thpb]
  \includegraphics[width=0.96\hsize]{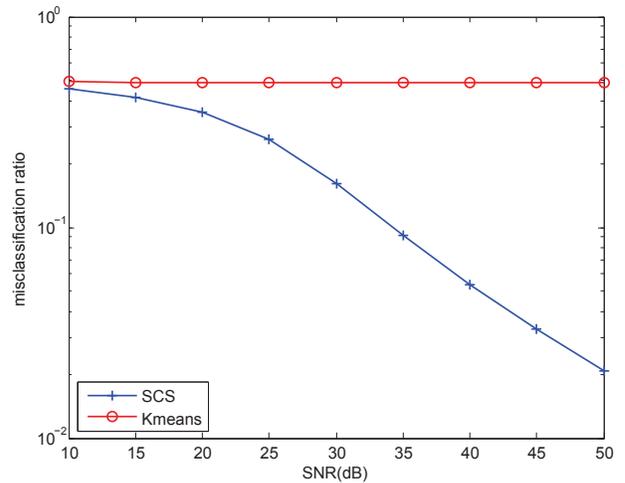}
  \caption{\label{fig_1D_mis} Misclassification ratio vs. SNR plots in Example 1.}
\end{figure}

The misclassification ratio is showed in Fig.~\ref{fig_1D_mis}. The $K$-mean based method in \cite{Ferrari} kept a constant misclassification ratio under all noise levels. SCS algorithm had misclassification error decays exponentially with respect to SNR.

\subsection{MIMO PLM with a chessboard domain partition\label{case2}}
\begin{figure}[thpb]
  \centering
  \includegraphics[width=0.96\hsize]{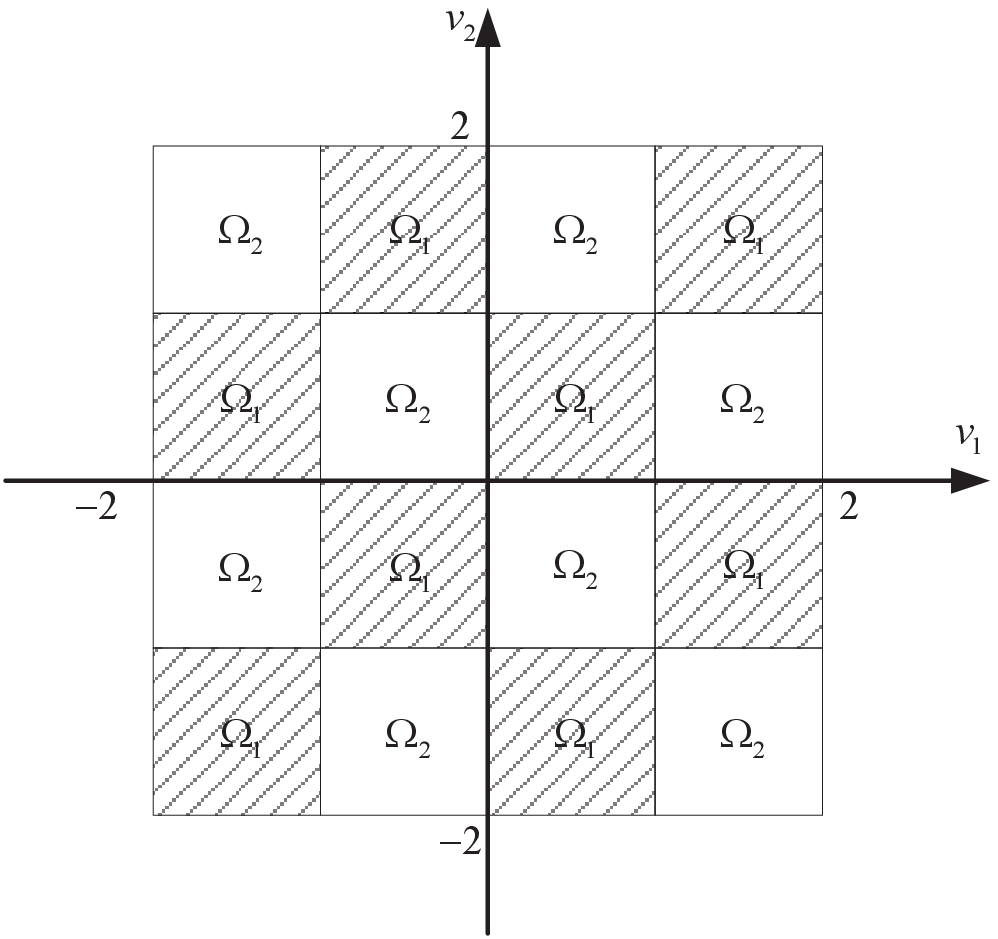}
  \caption{\label{fig_chess}The chessboard domain in 2-D case. Sub-model one is defined on the shadow areas $\Omega_1$; sub-model two is defined on the plain areas $\Omega_2$.}
\end{figure}

A main advantage of the proposed subspace approach is that the input domain partition can be arbitrarily shaped
and it deals with MIMO system with relatively easily.  In this simulation, we consider an example where the input domain partition is a ``chessboard'' where $\Omega_i$s are not connected.  See  Fig.~\ref{fig_chess}.

The system equations are given by
\begin{equation}
\begin{split}
  y_n&=\left\{\begin{array}{ll}
    \left[\begin{array}{cc}
      0.7&0.4\\
      0.5&0.3
    \end{array}\right]d_n+w_n&d_n\in\Omega_1\\
    & \\
    \left[\begin{array}{cc}
      0.8&0.9\\
      0.2&0.5
    \end{array}\right]d_n+w_n&d_n\in\Omega_2
  \end{array}\right.\\
  x_n&=\left[\begin{array}{cc}
      1&0\\
      0&1
    \end{array}\right]d_n+e_n.
\end{split}
\end{equation}
In the simulations, the input sequences were generated uniformly.

\begin{figure*}[thpb]
  \centering
  \subfigure[MSE of $\Theta_1(1,1)$]{\label{fig:fig_2D_mse:a}\includegraphics[width=0.48\hsize]{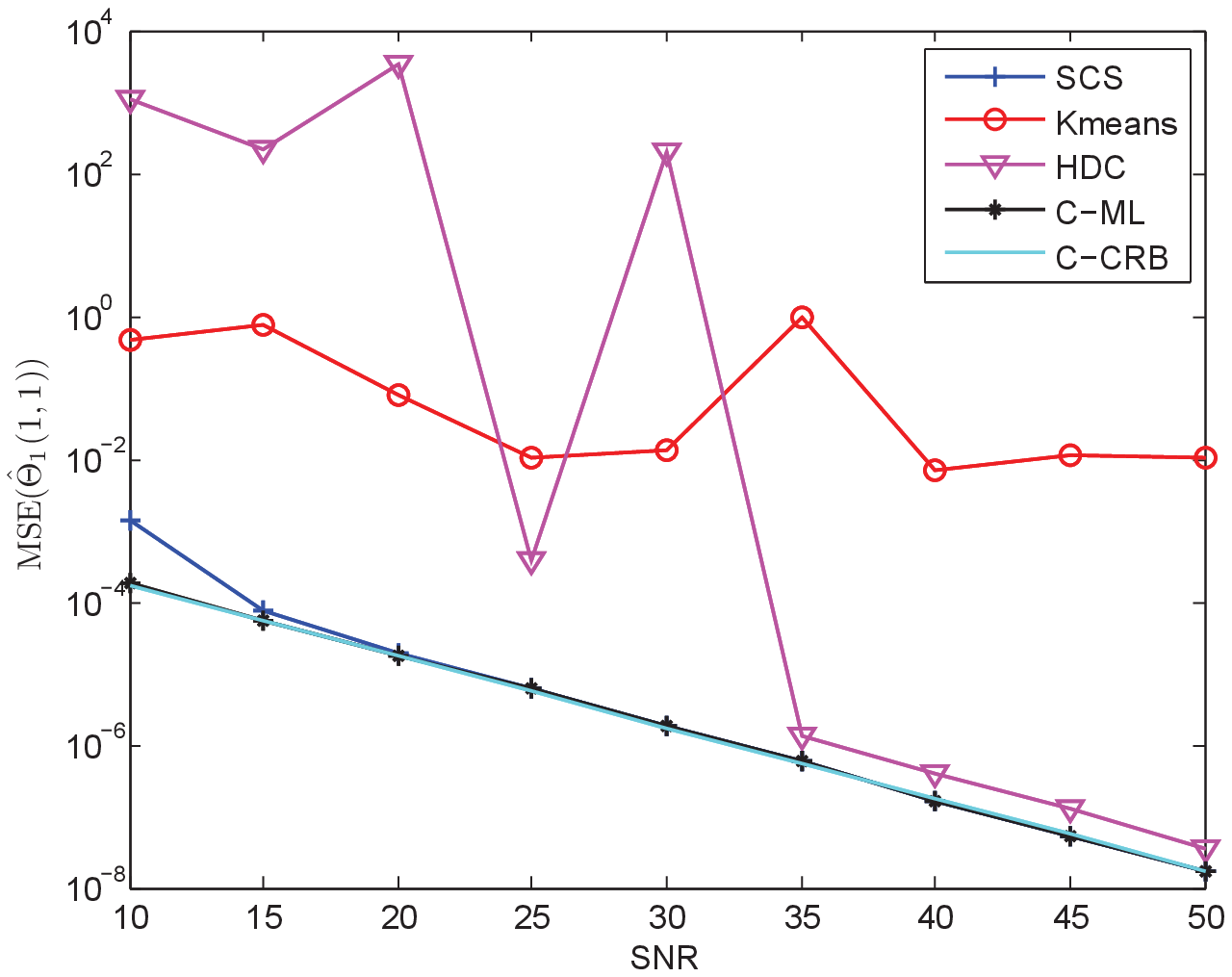}}
  \subfigure[MSE of $\Theta_2(1,1)$]{\label{fig:fig_2D_mse:b}\includegraphics[width=0.48\hsize]{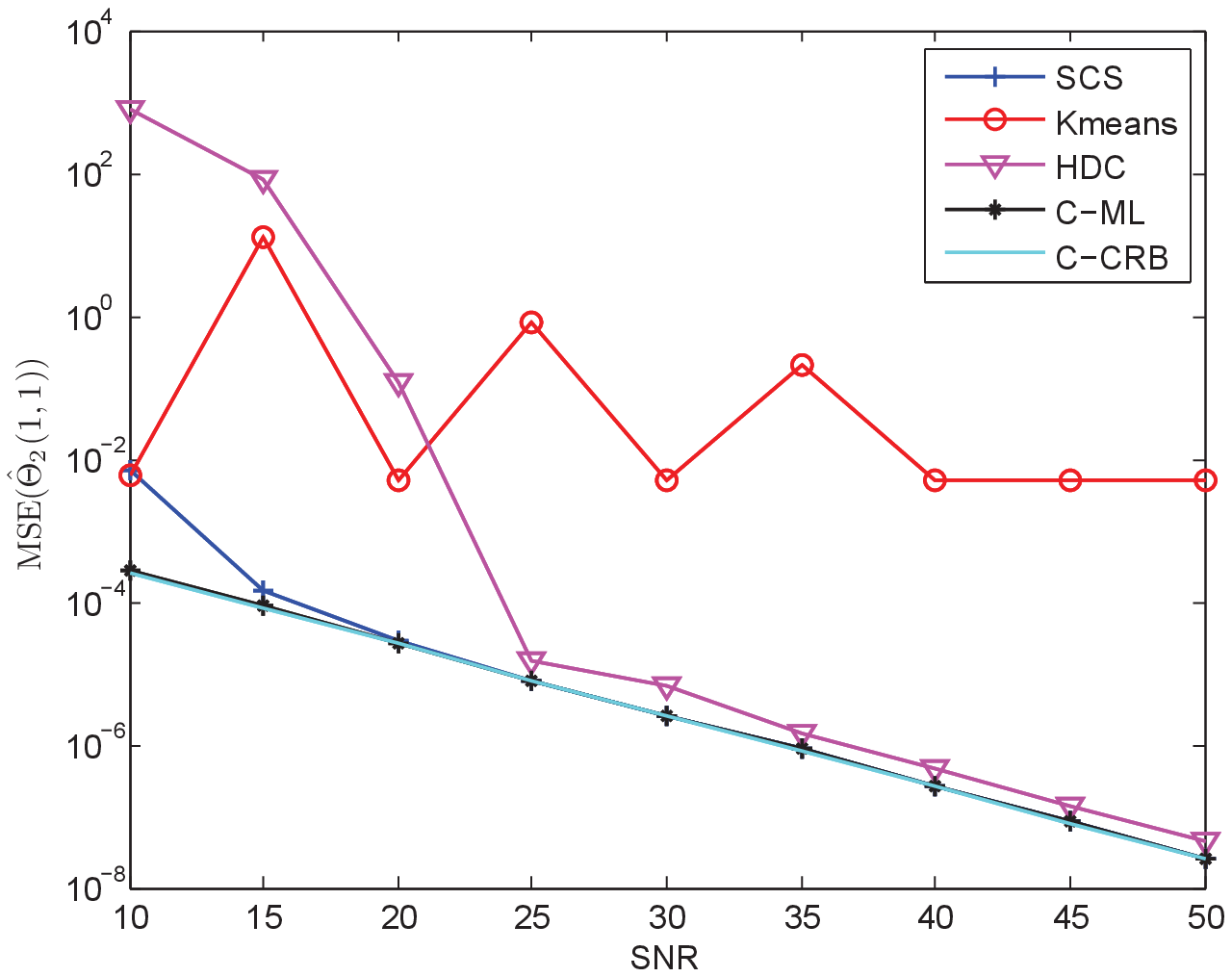}}
  \subfigure[Average MSE of $\Theta_1$]{\label{fig:fig_2D_mse:c}\includegraphics[width=0.48\hsize]{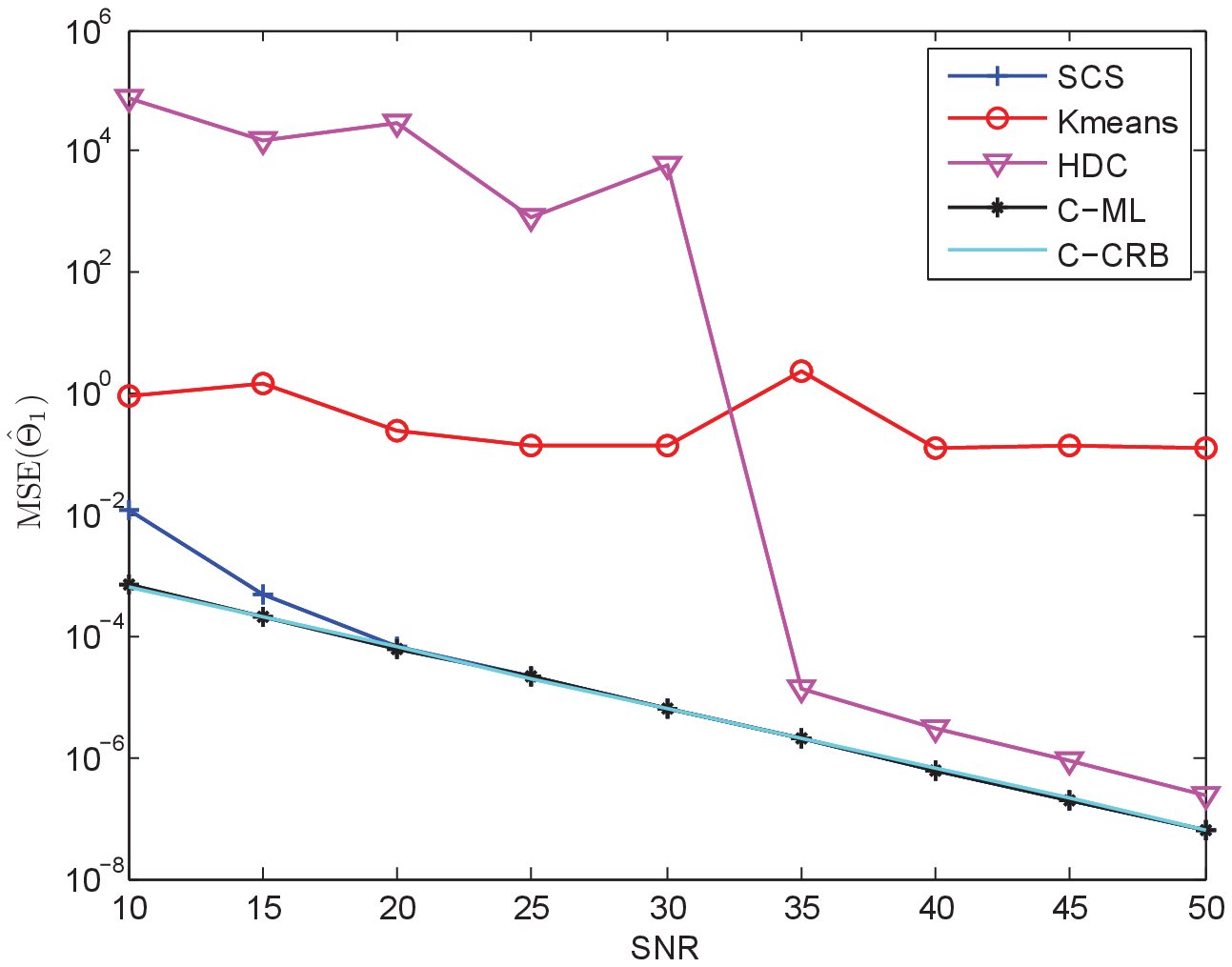}}
  \subfigure[Average MSE of $\Theta_2$]{\label{fig:fig_2D_mse:d}\includegraphics[width=0.48\hsize]{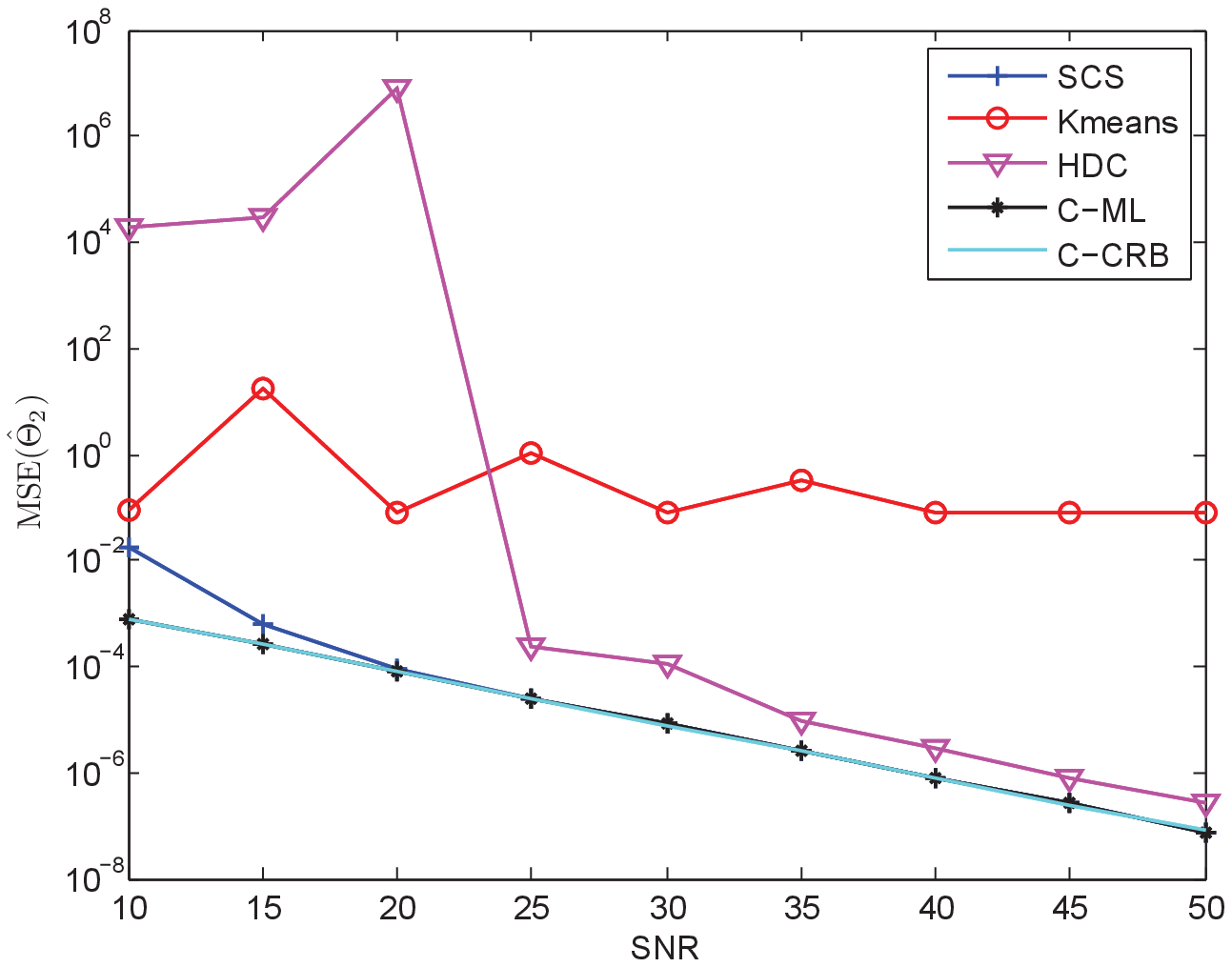}}
  \caption{\label{fig_2D_mse}Estimated MSE of $\Theta_1$ and $\Theta_2$ in chessboard domain case. Average MSE is the mean of all entries of parameter matrix. The data amount are $N_1=100\times8$ and $N_2=100\times8$, Monte Carlo runs is $10^3$ . SCS: the spectral clustering on subspace algorithm; Kmeans: the $K$-mean based method in \cite{Ferrari}, the local data set number is chosen as $c=5$; HDC: algebraic method based on \textit{hybrid decoupled constraints} \cite{Vidal}.
  C-ML: the maximum likelihood solution with the labels of observation known; C-CRB: the ``clairvoyant'' Cram\'{e}r-Rao bound.}
\end{figure*}

The MSE performance comparisons are shown in Fig.~\ref{fig_2D_mse}.  The proposed subspace estimator did not suffer the non-convex domain partition and had similar MSE performance as in the SISO case, matching the ``clairvoyant'' Cram\'{e}r-Rao bound when the SNR was greater than 20dB. As in many such subspace methods, the proposed algorithm was shown to have an SNR threshold such that the performance significantly improves after the SNR exceeds such a threshold.

The disconnected input domain partition presents a challenging scenario for the K-mean based algorithm.  In the simulation, the MSE performance of the $K$-mean algorithm did not improve beyond SNR=35dB. An explanation is that the ``chessboard'' domain partition generates output data overlapping in the observation space. The MSE performance of the $K$-mean technique saturates even as $\mbox{SNR}\rightarrow \infty$.

The Algebraic technique does give satisfied identification in the absence of observation noise.  However, we observed a fluctuated MSE in low SNR, due to the sensitivity of solutions of higher order polynomial equations. At high SNRs, algebraic geometry technique kept a constant gap from the ``clairvoyant'' Cram\'{e}r-Rao bound  as SNR increases, see Fig.~\ref{fig:fig_2D_mse:c}-\ref{fig:fig_2D_mse:d}, and its bias would be further amplified because the more parameters were involved in MIMO case.

\begin{figure}[thpb]
  \centering
  \includegraphics[width=0.96\hsize]{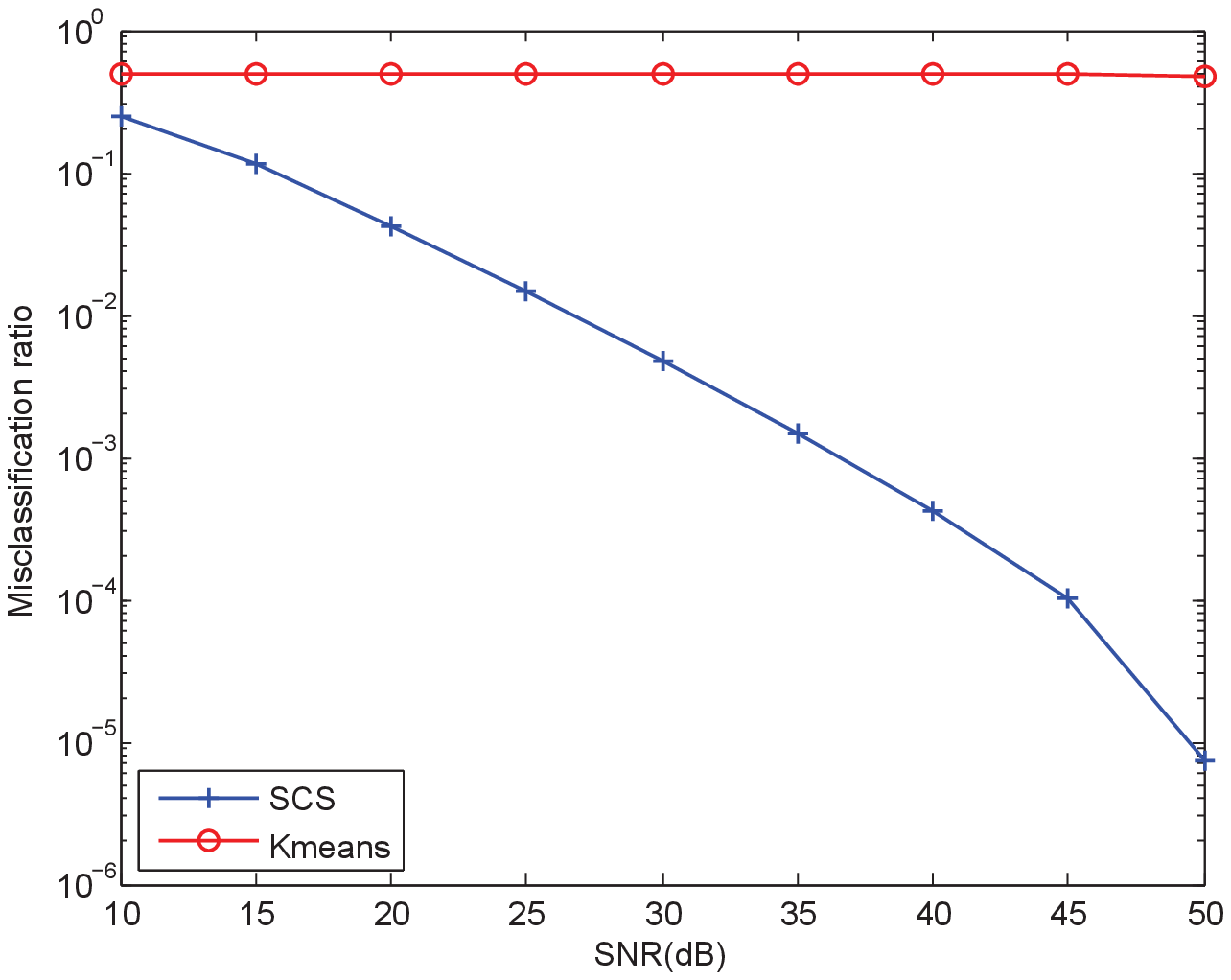}
  \caption{\label{fig_2D_mis}Misclassification ratio vs. SNR plots in chessboard domain case.}
\end{figure}

The misclassification ratio vs. SNR plot for ``chessboard'' domain is shown in Fig.~\ref{fig_2D_mis}. Similar curves  were achieved as in SISO case.
